\documentclass[]{article}
\makeatletter\if@twocolumn\PassOptionsToPackage{switch}{lineno}\else\fi\makeatother

\usepackage{amsmath,amsfonts,amsbsy,amssymb,tabulary,graphicx,times,caption,fancyhdr,amsthm, rotating,enumerate,bm}
\usepackage[utf8]{inputenc}
\usepackage[paperheight=10in,paperwidth=6.5in,margin=2cm,headsep=.5cm,top=2cm,headheight=1cm]{geometry}
\renewenvironment{abstract} {\vspace*{-1pc}\trivlist\item[]\leftskip\oupIndent\hrulefill\par\vskip4pt\noindent\textbf{\abstractname}\mbox{\null}\\\relax}{\par\noindent\hrulefill\endtrivlist} 
 \date{}
\usepackage{caption,booktabs}

\makeatletter\def\oupIndent{1pt}
\def\author#1{\gdef\@author{\hskip-\dimexpr(\tabcolsep)\hskip\oupIndent\parbox{\dimexpr\textwidth-\oupIndent}{\centering\bfseries#1}}}
\def\title#1{\gdef\@title{\centering\bfseries\ifx\@articleType\@empty\else\@articleType\\\fi#1}}
\let\@articleType\@empty \def\articletype#1{\gdef\@articleType{{\normalfont\itshape#1}}}
\fancypagestyle{headings}{\fancyhf{}\fancyhead[C]{\RunningHead}\fancyhead[R]{\thepage}}
\makeatother

\usepackage{url,multirow,morefloats,floatflt,cancel,tfrupee}
\makeatletter

\AtBeginDocument{\@ifpackageloaded{textcomp}{}{\usepackage{textcomp}}}
\makeatother
\usepackage{colortbl}
\usepackage{xcolor}
\usepackage{pifont}
\usepackage[nointegrals]{wasysym}
\makeatletter

\def\mcWidth#1{\csname TY@F#1\endcsname+\tabcolsep}

\def\cAlignHack{\rightskip\@flushglue\leftskip\@flushglue\parindent\z@\parfillskip\z@skip}
\def\rAlignHack{\rightskip\z@skip\leftskip\@flushglue \parindent\z@\parfillskip\z@skip}

\@ifundefined{etal}{}{}

\usepackage{ifxetex}
\ifxetex\else\if@twocolumn\@ifpackageloaded{stfloats}{}{\usepackage{dblfloatfix}}\fi\fi

\AtBeginDocument{
\expandafter\ifx\csname eqalign\endcsname\relax
\def\eqalign#1{\null\vcenter{\def\\{\cr}\openup\jot\m@th
  \ialign{\strut$\displaystyle{##}$\hfil&$\displaystyle{{}##}$\hfil
      \crcr#1\crcr}}\,}
\fi
}

\AtBeginDocument{%
  \@ifpackageloaded{endfloat}%
   {\renewcommand\efloat@iwrite[1]{\immediate\expandafter\protected@write\csname efloat@post#1\endcsname{}}}{\newif\ifefloat@tables}%
}%

\def\BreakURLText#1{\@tfor\brk@tempa:=#1\do{\brk@tempa\hskip0pt}}
\let\lt=<
\let\gt=>
\def\processVert{\ifmmode|\else\textbar\fi}

\@ifundefined{subparagraph}{
\def\subparagraph{\@startsection{paragraph}{5}{2\parindent}{0ex plus 0.1ex minus 0.1ex}%
{0ex}{\normalfont\small\itshape}}%
}{}

\newcommand\role[1]{\unskip}
\newcommand\aucollab[1]{\unskip}

\@ifundefined{tsGraphicsScaleX}{\gdef\tsGraphicsScaleX{1}}{}
\@ifundefined{tsGraphicsScaleY}{\gdef\tsGraphicsScaleY{.9}}{}
\def\checkGraphicsWidth{\ifdim\Gin@nat@width>\linewidth
	\tsGraphicsScaleX\linewidth\else\Gin@nat@width\fi}

\def\checkGraphicsHeight{\ifdim\Gin@nat@height>.9\textheight
	\tsGraphicsScaleY\textheight\else\Gin@nat@height\fi}

\def\fixFloatSize#1{}
\let\ts@includegraphics\includegraphics

\def\inlinegraphic[#1]#2{{\edef\@tempa{#1}\edef\baseline@shift{\ifx\@tempa\@empty0\else#1\fi}\edef\tempZ{\the\numexpr(\numexpr(\baseline@shift*\f@size/100))}\protect\raisebox{\tempZ pt}{\ts@includegraphics{#2}}}}

\AtBeginDocument{\def\includegraphics{\@ifnextchar[{\ts@includegraphics}{\ts@includegraphics[width=\checkGraphicsWidth,height=\checkGraphicsHeight,keepaspectratio]}}}

\DeclareMathAlphabet{\mathpzc}{OT1}{pzc}{m}{it}

\def\URL#1#2{\@ifundefined{href}{#2}{\href{#1}{#2}}}

\def\UrlOrds{\do\*\do\-\do\~\do\'\do\"\do\-}%
\g@addto@macro{\UrlBreaks}{\UrlOrds}

\edef\fntEncoding{\f@encoding}

\makeatother

\newif\ifmultipleabstract\multipleabstractfalse%

\makeatletter
\newcommand*{\centerfloat}{%
  \parindent \z@
  \leftskip \z@ \@plus 1fil \@minus \textwidth
  \rightskip\leftskip
  \parfillskip \z@skip}
\makeatother

\AtBeginDocument{\let\citet\cite\let\citep\cite}
\usepackage[numbers,sort&compress,numbers,sort&compress]{natbib}

\usepackage[T1]{fontenc}

\makeatletter
\renewcommand{\boxed}[1]{\text{\fboxsep=.2em\fbox{\m@th$\displaystyle#1$}}}
\makeatother



\usepackage[flushleft]{threeparttablex}
\usepackage{setspace,lscape}
\usepackage{algorithm,algpseudocode}
\usepackage{algpseudocode}
\usepackage{yhmath}
\usepackage{float}
\usepackage{enumitem}
\setlist[enumerate]{itemsep=0mm,leftmargin=5mm}
\begin{document}

\nocite{*}

\title{On the use of auxiliary variables in multiple imputation when estimating the average causal effect with missing data}

\author{\textbf{\fontsize{10pt}{16.4pt}\selectfont{Jiaxin Zhang\textsuperscript{1,2}, S. Ghazaleh Dashti\textsuperscript{1,2}, John B. Carlin\textsuperscript{1,2},  Katherine J. Lee\textsuperscript{1,2} and Margarita Moreno-Betancur\textsuperscript{2,1,*}}}~\\
{\small\normalfont \textsuperscript{1} Clinical Epidemiology and Biostatistics Unit, Murdoch Children’s Research Institute, Australia \\
\textsuperscript{2} Clinical Epidemiology and Biostatistics Unit, Department of Paediatrics, University of Melbourne, Australia\\
\textsuperscript{*}  jiaxin.zhang@mcri.edu.au~}}
\def\RunningHead{{}}

\maketitle 
\def\keywordstitle{Wordcount}

\abstract{
Estimating the average causal effect (ACE) using observational data is a key focus in causal inference for which missing data present an important challenge. Multiple imputation (MI) is a widely used method for handling missing data and can yield unbiased estimates when the imputation is compatible with the substantive analysis. One of the advantages of MI is its scope to include so-called ``auxiliary variables’’, defined as variables associated with incomplete variables that are excluded from the substantive analysis. Although many studies have looked at the use of auxiliary variables in MI for improving precision, the study of auxiliary variables that are necessary for the identifiability (or “recoverability”) of the ACE in the presence of missing data has been scant. In this work, we investigate the use of auxiliary variables, both mediators and non-mediators, across a range of typical univariable and multivariable missingness mechanisms depicted by missingness directed acyclic graphs (m-DAGs). For each setting, we derive recoverability results, then evaluate MI-based and complete-case methods for estimating the ACE using correctly specified g-computation, considering different strategies for incorporating auxiliary variables and varying degrees of compatibility for MI models. Based on findings from the simulation studies, we provide practical guidance, highlighting that distinguishing appropriately between mediator and non-mediator auxiliary variables is important to avoid bias as is the use of compatible and flexible (non-parametric) MI methods that incorporate these variables.
} 

\def\keywordstitle{Keywords}

\def\keywordstitle{Keymessages}

\clearpage
\section{Introduction}\label{sec: Introduction}
In epidemiological studies addressing causal questions, the average causal effect (ACE) is a common target estimand. Formally, the ACE is defined as the contrast between the average potential outcome under exposure and no exposure \cite{Causal2020}. In the absence of measurement error, missing data and other sources of selection bias, the ACE is non-parametrically identifiable under well-known assumptions (consistency, positivity, and exchangeability given confounders), meaning it can be expressed as a function of observable data in an infinite sample using the g-formula, and can thus be unbiasedly estimated using a range of methods. One such method is g-computation, which requires a correctly specified outcome model \cite{robins1986new,snowden2011implementation}.

In the presence of missing data, additional assumptions about the missingness mechanism are required for non-parametric identifiability (also referred to as recoverability in the missing data context) of the ACE. As above, the ACE is recoverable if, under the assumed missingness mechanism and other causal assumptions, it can be mathematically expressed as a function of the distribution of the observable data in an infinite sample \cite{mohan2014graphical}. Several studies have investigated the recoverability of the ACE under typical multivariable missingness mechanisms depicted by missingness directed acyclic graphs (m-DAGs), which represent the assumed causal structure among variables relevant to the analysis and variable-specific missingness indicators \cite{moreno2018canonical,zhang2024recoverability,zuo2025mediation,moreno2025correction}. These studies have shown that, in some scenarios, recoverability expressions can guide the choice of an appropriate method for handling missing data \cite{lee2023assumptions}. For instance, if the ACE is recoverable and can be written solely as a function of distributions restricted to the complete cases (i.e. records without missing values in analysis variables), then a complete-case analysis (CCA) enables consistent estimation of the ACE. Motivated by the widespread use of multiple imputation (MI) for handling missing data \cite{ rubin2004multiple,carpenter2023multiple}, a subset of these studies has also examined the performance of MI across m-DAGs using simulations \cite{moreno2018canonical,zhang2024recoverability,dashti2024handling,moreno2025correction}, providing insights into its bias across different causal structures, as well as examining potential gains in precision. 

MI is a widely used approach to handling missing data \cite{rubin2004multiple,white2010bias,little2019statistical,carpenter2023multiple}. An often-cited advantage of MI over CCA is that it can incorporate information from variables not included in the substantive analysis but associated with the incomplete variable and possibly also predictive of missingness, i.e. auxiliary variables. The traditional MI literature recommends including auxiliary variables to improve precision and potentially reduce bias. From a modern causal perspective, bias reduction when including auxiliary variables in MI would be expected in settings where the auxiliary variables are necessary for establishing recoverability of the target estimand, in the sense that they are required to block problematic open paths between an incomplete variable and its missingness indicators \cite{thoemmes2014cautious,curnow2023multiple,mathur2024imputation,curnow2024multiple}. However, existing studies investigating the recoverability of the ACE under typical multivariable missingness mechanisms, and associated performance of MI, have not examined settings with such auxiliary variables. Consideration of auxiliary variables required for recoverability raises two key challenges, namely the choice of missing data method and, when MI is used, its implementation. 

The first challenge, relating to the choice of method, arises when the auxiliary variables in question are mediators on the path from exposure to outcome. This scenario is common in longitudinal studies, where intermediate outcomes affected by exposure serve as auxiliary variables, as they are common causes of later outcomes and their missingness indicators. Including these variables somehow in the missing data approach may be necessary to block problematic open paths, but some approaches will not work. For example, a line of work has examined a form of CCA, referred to as the adjusted-CCA (A-CCA) approach, that includes non-mediator auxiliary variables in an expanded adjustment set in an outcome regression model \cite{mathur2025common,mathur2024imputation}). However, such an approach would not be appropriate for mediator auxiliary variables when the target estimand is the ACE, because these variables lie on the causal pathway. In this case, MI appears to be a promising alternative, but this has yet to be examined \cite{thoemmes2014cautious,curnow2023multiple,mathur2024imputation,curnow2024multiple}. 

The second challenge if MI is the method of choice, is how to include auxiliary variables in MI (whether mediators or not) while ensuring compatibility between the imputation model and the substantive analysis model; otherwise, MI estimates may be biased \cite{meng1994multiple,bartlett2015multiple}. Briefly, compatibility means that the parametric assumptions made in the imputation model do not conflict with those made in the analysis model. For example, when estimating the ACE using g-computation with a correctly specified outcome model, any interactions included in the outcome model should also be incorporated into the imputation model to avoid incompatibility \cite{zhang2024recoverability}. Another form of incompatibility arises when the imputation model and the analysis model include different sets of variables, such as when auxiliary variables are used in MI, as these variables are by definition not in the analysis model \cite{carpenter2023multiple}. Although different implementations of MI have been proposed to address incompatibility in the presence of auxiliary variables \cite{carpenter2023multiple}, it remains unclear how MI should be implemented and how the different implementations perform when estimating the ACE in multivariable missingness settings where auxiliary variables, including potential mediators, are required to establish recoverability.

In this work, we aimed to address the gaps described above relating to the choice and implementation of missing data methods in settings where auxiliary variables are required for recoverability of the ACE. The paper is organised as follows. First, we introduce a motivating example from the Victorian Adolescent Health Cohort Study (VAHCS) in section \ref{sec: Motivating example}. We then define the notation and the ACE estimand, followed by a detailed review of existing recoverability results and MI approaches considered in this paper in section \ref{sec: Preliminaries}. In Section \ref{sec: Methods}, we describe a set of m-DAGs that represent typical univariable and multivariable missingness mechanisms with mediator and non-mediator auxiliary variables required for recoverability, and provide recoverability expressions for the ACE in each m-DAG. In Section \ref{sec: Simulation}, we describe the simulation study we conducted to evaluate and compare the performance of various missing data methods across the m-DAGs described, which are CCA, A-CCA and different implementations of MI with auxiliary variables. In Section \ref{sec: Case}, we describe results from the VAHCS case study. Finally, in Section \ref{sec: Discussion}, we summarise and discuss our findings.

\section{Motivating example}\label{sec: Motivating example}
The motivating example we use for illustration and to inform the design of our simulation study is based on an investigation by Patton et al \cite{patton2002cannabis} examining the causal effect of frequent cannabis use in female adolescents on their risk of mental health problems in young adulthood. The study used data from the Victorian Adolescent Health Cohort Study (VAHCS), a longitudinal cohort study that recruited 1943 students ($n$=1000 females) aged 14 to 15 years from schools in Victoria, Australia, in 1992-1993. Participants were surveyed for health and behaviour measurements every six months in their adolescence (waves 2 to 6). Outcome was measured in Wave 7 when participants were in early adulthood (age 20 years). 

\begin{figure}[h!]
\centering
\includegraphics[width=0.8\linewidth]{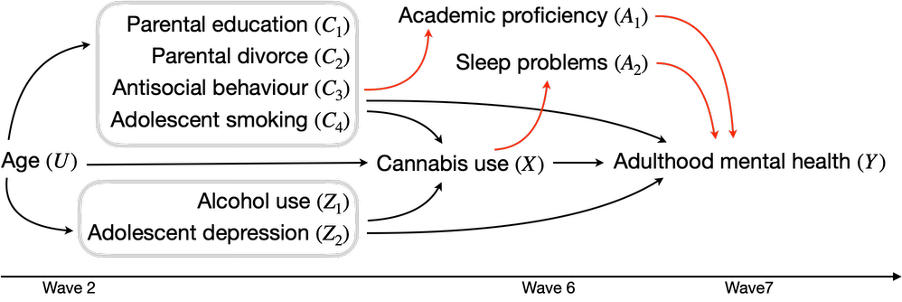}
\caption{Directed acyclic graph (DAG) used to guide data generation for the simulation study, based on a motivating example from the Victorian Adolescent Health Cohort Study (VAHCS)}
\label{fig: V-DAG}
\end{figure}

Following the published study \cite{patton2002cannabis}, we treated the exposure as a single time-point measure, defined as reporting cannabis use more than once a week in any of the adolescent waves. The outcome was the mental health score at wave 7, measured using the $z$-score of the raw computerised Clinical Interview Schedule – Revised (CIS-R) score \cite{lewis1992manual}. We included all the confounders in the original study \cite{patton2002cannabis}, including parental education and parental divorce by wave 6, as well as antisocial behaviour, adolescent depression, and alcohol use across adolescent waves. We additionally included adolescent smoking as a confounder, following recent literature \cite{goriounova2012short,coffey2016cannabis,colyer2023age}. The confounders were grouped according to their missingness proportions: confounders $C_1-C_4$ were (nearly) complete, whereas confounders $Z_1$ (alcohol use) and $Z_2$ (adolescent depression) were incomplete. In depicting the assumed causal structure (Figure \ref{fig: V-DAG}), we included $U$, participant’s age at wave 2 ($z$-score), as an auxiliary variable representing a common cause of exposure and confounders -- this auxiliary variable is not required for recoverability but could improve precision in MI. The participant’s age has 9.3\% missing values in the VAHCS case study. Given focus on auxiliary variables required for recoverability, throughout the manuscript (except for the case study application) we used $U$ to model an unobserved factor inducing correlations amongst exposure and confounders. Additionally, we included two additional auxiliary variables required for recoverability (as shown later), which are the focus of the paper: $A_1$, academic proficiency ($z$-score) in adolescence (waves 2 to 6), and $A_2$, sleep problems at wave 6. Both are assumed to be causes of the outcome in this example, with $A_2$ additionally acting as a mediator between the exposure and the outcome. We acknowledge that the resulting DAG (Figure \ref{fig: V-DAG}) may be debatable from a substantive viewpoint, but we focus on this simplified structure to enable examination of the two types of auxiliary variables of interest (mediator and non-mediator). Table \ref{tab: table1} presents descriptive statistics and missing data proportions for each variable in the VAHCS data. 

\section{Preliminaries and overview of existing methods}\label{sec: Preliminaries}
\subsection{Notation}\label{subsec: Notation}
Similar to the set-up for the motivating example (Figure \ref{fig: V-DAG}), let $X$, $Y$, $\bm{C}$ and $\bm{Z}$ denote the exposure, outcome, set of complete confounders and set of incomplete confounders, respectively, where $X$ is binary (coded 0/1) but the other variables can be of any type. Let $U$ denote an unmeasured common cause of exposure and confounders. Let $Y^x$ denote the potential outcome under exposure $X=x$, and $P(Y^x=y)$ represent the probability distribution of the potential outcome. We use the shorthand $P(a)$ to represent $P(A=a)$, and in particular $P(y^x)$ to represent $P(Y^x=y)$, and we also use $P(a|b)$ as shorthand for $P(A=a|B=b)$. Let $\delta$ denote the target estimand, which is the ACE as defined in the next section. 

We consider settings where the exposure, outcome and some confounders ($Z$) are subject to missingness. Let $M$ with a subscript denote the missingness indicator of the incomplete variable in the subscript, e.g. $M_Y=1$ if $Y$ is missing and $M_Y=0$ if $Y$ is observed. Let $M_{all}=0$ denote a missingness indicator where $M_{all}=0$ if and only if $M_X=M_Y=M_{\bm{Z}}=0$, and $M_{all}=1$ otherwise; thus, conditioning an analysis on $M_{all}=0$ equates to restricting the analysis to fully observed records, i.e. a CCA. In considering different m-DAGs, we use $A_1$ and $A_2$ to denote completely observed auxiliary variables that are causes of the outcome and the missingness indicators for the exposure ($M_X$) and/or the outcome ($M_Y$) and such that $A_2$ (but not $A_1$) is caused by the exposure. Let $W$ an unmeasured common cause of missingness indicators. 

\subsection{Estimand, identifiability and estimation in the absence of missing data}\label{subsec: Estimand}
\begin{figure}[h!]
\centering
\includegraphics[width=0.7\linewidth]{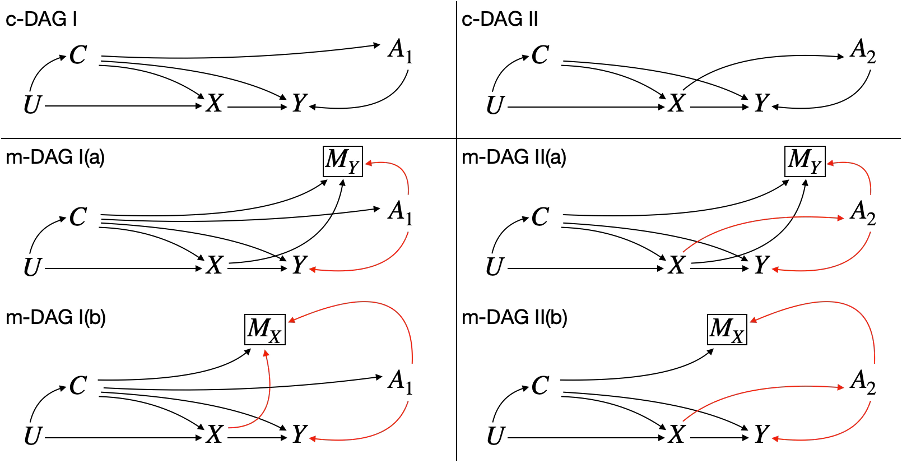}
\caption{DAGs depicting typical settings for auxiliary variables in univariable missingness mechanisms. On the left, c-DAG I, and m-DAGs I(a) and I(b) show settings where the auxiliary variable \emph{is not caused} by the exposure (i.e. it is not a mediator). On the right, c-DAG II, and m-DAGs II(a) and II(b) show settings where the auxiliary variable \emph{is caused} by the exposure (i.e. it is a mediator). m-DAGs I(a) and II(a) depict incomplete outcome scenarios, and m-DAGs I(b) and II(b) depict incomplete exposure scenarios.}   
\label{fig: uni-DAG}
\end{figure}

The target estimand, the ACE, is defined as the difference in the expected potential outcomes if all participants were set to exposure versus no exposure, i.e. $\delta=E[Y^{x=1}]-E[Y^{x=0}]$. In the absence of missing data (i.e. complete exposure and outcome and $\bm{Z}$ being the empty set), the expected potential outcomes are identifiable from observable data by the g-formula under the assumptions of exchangeability given confounders $\bm{C}$, consistency, and positivity \cite{Causal2020}: 
\begin{equation}\label{equ: g-formula}
E[Y^x]=\int E(Y|X=x,\bm{C}=\bm{c}) dF_{\bm{C}}(\bm{c}).
\end{equation}
Without missing data, exchangeability assumptions can be portrayed using ``complete data’’ DAGs (c-DAGs) \cite{moreno2018canonical}. Figure \ref{fig: uni-DAG} shows two example c-DAGs where exchangeability given confounders $\bm{C}$ holds. Thus, assuming consistency and positivity, in both c-DAGs, $E[Y^x]$ is identifiable via the g-formula in (\ref{equ: g-formula}). 

Under the identifiability assumptions, several estimators are available for estimating the ACE from the data. Here, we focus on g-computation \cite{robins1986new,snowden2011implementation}. This approach relies on a correctly specified regression model for the outcome, including the exposure and confounders as regressors and possibly also exposure-confounder interaction terms. In the absence of missing data, a common form of this model is given by \cite{robins1986new,snowden2011implementation}: 
\begin{equation}\label{equ: outmod in g-comp}
E[Y|X,\bm{C}]=\alpha_0 + \alpha_1 X + \bm{C \alpha_2}^T + \bm{C} \bm{\alpha_3}^T X. 
\end{equation}
The ACE is estimated as the difference between the average of the outcomes predicted from the fitted conditional regression model (\ref{equ: outmod in g-comp}) when setting all individuals to be exposed versus unexposed \cite{robins1986new,snowden2011implementation}. 

\subsection{Identifiability with missing data}\label{subsec: Identifiability}
In the presence of missing data, additional assumptions about the missingness mechanism are required to determine whether the ACE is identifiable or, in terminology used in the missing data context, ``recoverable’’ \cite{mohan2014graphical}. For an assumed missingness mechanism, the marginal distribution of the potential outcome (and thus the ACE) is recoverable if it can be expressed as a function of distributions that either pertain to fully observed variables only, or, if they include an incomplete variable, they are conditioned on the corresponding missingness indicator being set to 0 \cite{mohan2014graphical,mohan2013graphical}. 

In this paper, we focus on settings where the recoverability expressions require conditioning on auxiliary variables. For example, under the assumed causal structure depicted in missingness-DAG (m-DAG) I (a) in Figure \ref{fig: uni-DAG} for a univariable missingness setting, the expected potential outcome is recoverable and can be expressed as:
\begin{equation*}
E[Y^x]=\int E(Y|X=x,A_1=a_1,\bm{C}=\bm{c},M_Y=0) dF_{A_1\bm{C}}(a_1,\bm{c}),
\end{equation*}
This equation indicates that estimating the ACE using g-computation with a correctly specified outcome model including the auxiliary variable $A_1$ (and possibly an exposure-auxiliary variable interaction) is unbiased. In settings such as that depicted by m-DAG I (a), where all other variables except for outcome are fully observed, recoverability of the expected potential outcome conditional on $A_1$ and $\bm{C}$ is sufficient for recoverability of $E[Y^x]$. In more complex scenarios, such as when confounders are incomplete, no such general condition has been established. In Section \ref{sec: Methods}, we provide recoverability results for a range of m-DAGs that reflect typical mechanisms where these results are currently lacking in practice. 

\subsection{MI to handle missing data}\label{subsec: Missing data methods}
MI is an approach for handling missing data, commonly consisting of the following two steps. First, missing values are imputed multiple times by drawing from an approximation to their posterior predictive distribution given the observed data (under an imputation model). Second, the substantive analysis (in this case, g-computation) is performed on each imputed dataset, and the resulting estimates and standard errors are combined using Rubin’s rules to produce a final estimate with an associated standard error. A flexible and widely used implementation of MI for handling multivariable missingness is fully conditional specification (FCS), also known as chained equations. In FCS, univariate imputation models are specified for each incomplete variable given other variables, and the algorithm cycles through these models iteratively, updating imputations until convergence.

When implementing MI, a key consideration for avoiding potential bias in estimates of interest is the compatibility between the imputation and substantive analysis models \cite{meng1994multiple,carpenter2023multiple}. These models are compatible if they correspond to conditional distributions for an overarching joint model \cite{liu2014stationary,bartlett2015multiple}. In real applications, it is often difficult to ensure strict compatibility. Therefore, the practical recommendation for reducing potential bias due to incompatibility is that the imputation model should be specified such that it accommodates all the characteristics of the substantive model and avoids making conflicting parametric assumptions \cite{tilling2016appropriate,zhang2024recoverability}. In the context of a substantive analysis using g-computation, this means that the imputation should accommodate the exposure-confounder interactions included in the outcome model. 

There are several approaches within FCS for accommodating interactions in the imputation model \cite{rubin2004multiple,van2007multiple}. Firstly, interactions can be handled by including the relevant interactions as predictors when using parametric regression-based (referred to as REG) \cite{van2007multiple,tilling2016appropriate} or semi-parametric predictive mean matching (PMM) \cite{white2011multiple} imputation approaches. Alternatively, imputation using a non-parametric approach like classification and regression trees (CART) automatically accommodates interactions \cite{doove2014recursive,breiman2017classification}. A more principled approach is a substantive model compatible FCS (SMCFCS) \cite{bartlett2015multiple}, which draws the imputations from a so-called proposal distribution that is proportional to the outcome distribution in a target model to ensure compatibility. This target model usually uses the same outcome model as that used for the substantive analysis, for example, the conditional regression model (\ref{equ: outmod in g-comp}) used in g-computation, but can be different (see next paragraph). Our previous simulation study has shown that under missingness mechanisms where the ACE is recoverable without using auxiliary variables, all of these MI approaches return an approximately unbiased estimate of the ACE \cite{zhang2024recoverability}. 

An MI approach that incorporates auxiliary variables cannot be strictly compatible with a substantive analysis model that does not include these variables. However, such incompatibility does not always result in bias if the imputation model is compatible with a larger model that contains the substantive analysis model, such that it keeps all features in the analysis model and additionally includes the auxiliary variables. In this case, the MI approach is said to be semi-compatible with the analysis model \cite{liu2014stationary,bartlett2015multiple}. When implementing the SMCFCS approach, there are two ways of incorporating auxiliary variables that ensure semi-compatibility \cite{carpenter2023multiple}. The first approach is to specify the target model as an extension of the outcome model by adding the auxiliary variables. Alternatively, the target model can be defined as the outcome model, while the auxiliary variables are included in the proposal distribution for drawing imputations. For other FCS approaches, i.e., REG and PMM, semi-compatibility can be achieved by including the auxiliary variables as predictors in the imputation model \cite{van2007multiple}. However, it is not clear how any of these approaches perform in realistic epidemiological settings when the auxiliary variables in question are necessary for establishing recoverability. We will examine this question in our simulation study (see Section \ref{subsec: implementation}).

\section{Missingness mechanisms with auxiliary variables}\label{sec: Methods}
In this section, we describe a range of univariable and multivariable missingness mechanisms in which auxiliary variables are necessary to establish the recoverability of the ACE, and we present recoverability results for the ACE under each mechanism. 

\subsection{Typical scenarios depicted by m-DAGs}\label{subsec: Typical}
\paragraph{Univariable missingness mechanisms}
Figure \ref{fig: uni-DAG} shows univariable missingness mechanisms in a simplified setting relative to the motivating example, with only a single confounder $C$, where either the outcome (m-DAGs I(a) and II(a)) or the exposure (m-DAGs I(b) and II(b)) is incomplete. In both cases, there is a single fully observed auxiliary variable that is a common cause of the outcome and the missingness indicator, and it may (m-DAGs II(a) and II(b)) or may not (m-DAGs I(a) and I(b)) also be caused by the exposure. As we will see in the next section, these m-DAGs represent settings where the auxiliary variable in question is necessary for establishing recoverability (and there are no other auxiliary variables required for recoverability). This means that if the auxiliary variable were unmeasured, then the ACE would not be recoverable. To see this in m-DAG I(a), $Y$ and $M_Y$ are neighbours and this path cannot be blocked when $A_1$ is unobserved, rendering the ACE nonrecoverable \cite{mohan2014graphical}. In m-DAG I(b), $Y$ and $X$ are connected by an open path through the collider $M_X$ that is inevitably conditioned upon, and again this path cannot be blocked when $A_1$ is unobserved, rendering the ACE non-recoverable. In both settings, standard methods such as CCA and MI are expected to be biased, as shown in previous simulation studies examining similar scenarios \cite{zhang2024recoverability}.  In m-DAGs II(a) and II(b), we extend these scenarios to settings where the auxiliary variable is a mediator, resulting in pattern of biasing paths if this variable were unobserved that are analogous. Note that the direct arrows from $X$ and $Y$ to $M_X$ are replaced by the paths through $A_2$ in m-DAG II(b), and it is important to notice that there is no way to recover ACE if this direct arrows are still present in the current m-DAG II(b). 

\begin{figure}[h!]
\centering
\includegraphics[width=\linewidth]{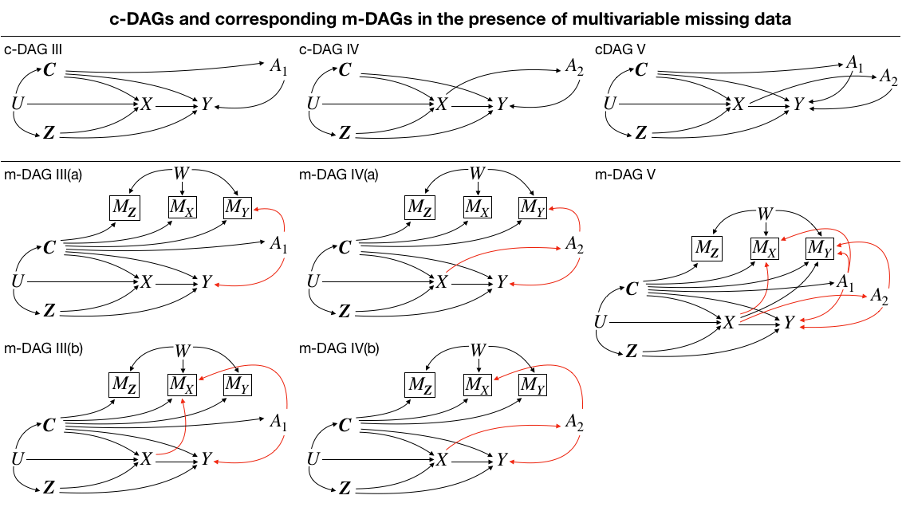}
\caption{DAGs depicting typical settings for auxiliary variables in multivariable missingness mechanisms.}
\label{fig: multi-DAG}
\end{figure}

\paragraph{Multivariable missingness mechanisms}
In real-world studies, it is common to encounter missing values in all analysis variables: the exposure, the outcome and at least some confounders. The m-DAGs in Figure \ref{fig: multi-DAG} depict a range of such settings. These m-DAGs extend those considered in \cite{moreno2018canonical,moreno2025correction,zhang2024recoverability} by including auxiliary variables that are required for establishing recoverability. These multivariable missingness mechanisms include an unmeasured common cause of the missingness indicators, denoted by $W$, to capture the correlation among missingness indicators that is typically present in practice. These m-DAGs can be viewed as extensions of the univariable m-DAGs in Figure \ref{fig: uni-DAG}, in that they depict similar relationships between the auxiliary and analysis variables. m-DAG V depicts a more general scenario that represents a plausible missingness mechanism in the VAHCS example, in which two auxiliary variables are present: $A_2$ is a mediator of the exposure-outcome relationship, and $A_1$ is not.
\subsection{Recoverability and estimation of the ACE}\label{subsec: Recoverability}
Table \ref{tab: recoverability} presents recoverability expressions for the marginal distribution of the potential outcome under a given exposure value, $P(y^x)$, for the m-DAGs shown in Figures \ref{fig: uni-DAG} and \ref{fig: multi-DAG}. These expressions are written for categorical variables but can be generalised to continuous variables by replacing summations over probabilities with integrals. Informally, for each m-DAG, the ACE is recoverable if the auxiliary variable is fully observed, since conditioning on it blocks the red path between the missingness indicator (for exposure or outcome) and the outcome. 

The ACE can in principle be unbiasedly estimated from the exact recoverability expressions by modelling all distributions, but this would be highly impractical: not only are some of these expressions highly complex, relying on modelling multiple distributions, but the specific expression and corresponding estimation method would need to be derived and implement for each problem individually. Additionally, all this work would be based on an assuming the m-DAG holds but the m-DAG is inevitably subject to uncertainty in most settings \cite{vanderweele2019principles}. Therefore, a more practical approach is to consider flexible and accessible methods that can incorporate auxiliary variables that are required for recoverability and are likely to provide approximately unbiased estimation across a wide range of applications. We next describe some potential approaches and evaluate them in the next section.

One intuitive and easy way to incorporate auxiliary variables that are not mediators, is to include them as covariates in the outcome model. We refer to the model that adjusts for both confounders and auxiliary variables as the A-adjusted model, while the model that only adjusts for confounders, but not auxiliary variables, is referred to as the C-adjusted model, i.e. the conditional outcome model (\ref{equ: outmod in g-comp}). For example, in m-DAG I(a) (and similarly in m-DAG I(b)), one possible specification of an A-adjusted model would include $A_1$ as a main effect:  
\begin{equation}\label{equ: A-model}
E[Y|X,A_1,\bm{C}]=\beta_0 + \beta_1 X + \beta_2A_1 + \bm{C \beta_3}^T + \bm{C} \bm{\beta_4}^T X. 
\end{equation}
Accordingly, analogous to the CCA approach for estimating the ACE using regular g-computation (with a C-adjusted model, referred to as C-CCA), one can conduct g-computation with an A-adjusted model on the complete cases (referred to as A-CCA). 

Alternatively, and necessarily when auxiliary variables are mediators, they can be incorporated in the MI procedure when imputing the missing values, followed by g-computation with the C-adjusted model using the imputed datasets. In the next section, we describe a simulation study we performed to evaluate the performance of different methods for handling missing data across the m-DAGs in Figures \ref{fig: uni-DAG} and \ref{fig: multi-DAG}.

\section{Simulation study}\label{sec: Simulation}
In this section, we describe a simulation study for evaluating the performance of methods for handling missing data described in Sections \ref{subsec: Missing data methods} and \ref{subsec: Recoverability}, across the univariable and multivariable missingness settings depicted by the m-DAGs in Figures \ref{fig: uni-DAG} and \ref{fig: multi-DAG}. To maintain consistency with the example, we considered a continuous outcome in all scenarios and defined the ACE on the difference in means scale as $\delta=E[Y^{x=1}]-E[Y^{x=0}]$. 

\subsection{Aims}
The aim of our simulation was two-fold, motivated by the fact that, as mentioned above, the methods evaluated are in general not consistent (i.e. not based on the recoverability expression) and thus expected to yield some bias.\begin{itemize}
\item Our first aim was to examine the presence and extent of bias of methods in simplified and extreme (in terms of strength of associations) univariable missingness scenarios.
\item Our second aim was to examine overall method performance, across a range of indicators, under realistic scenarios with multivariable missingness.
\end{itemize}

\subsection{Data generation}
\paragraph{The univariable missingness mechanisms under strong associations} We simulated data under the four missingness mechanisms depicted by m-DAGs I(a), I(b), II(a) and II(b) in Figure \ref{fig: uni-DAG}. We generated $U$ from a standard Normal distribution and generated a binary complete confounder $C$ and the binary exposure $X$ sequentially from Bernoulli distributions: $C \sim Bern(\text{expit}(0.5+U))$ and $X \sim Bern(\text{expit}(-0.5+C+U))$. This resulted in an exposure prevalence of around 50\%. For each m-DAG, we considered two separate scenarios differing in the type of the auxiliary variable (binary or continuous; see generation models in Table \ref{tab: aux-simu}) and two outcome generation scenarios that differed by whether an exposure–confounder interaction was included or not. Overall, we considered 16 scenarios across the m-DAGs and auxiliary variable and outcome generation scenarios. Outcomes were generated using normal linear models with the following mean specifications, where the auxiliary variable is considered to have strong effect on the outcome:
\begin{equation}\label{equ: A-mod uni}
\begin{aligned}
E[Y|X,A_k,C]&= X+2A_k+C & \text{(no interaction scenario)} & \\
E[Y|X,A_k,C]&= X+2A_k+C+XC & \text{(interaction scenario)}&, \\
\end{aligned}
\end{equation}
where $k=1$ for m-DAGs I(a) and I(b) and $k=2$ for m-DAG II. The missingness indicator ($M_X$ or $M_Y$) was generated from a Bernoulli distribution $Bern(\text{expit}(\eta_0+\eta_1X-C+A_k))$, where $\eta_1=0$ for m-DAG II(b) and $\eta_1=1$ for other m-DAGs; $\eta_0$ was modified in order to fix the missingness proportion at 35\%. The true value of the ACE in each scenario is marked by a dashed line in the ``Estimate’’ panel of Figure \ref{fig: uni-res}.

\paragraph{The multivariable missingness mechanisms under realistic associations} The parameter values used for the data generation models were estimated by fitting analogous models to the VAHCS data unless stated otherwise. Parameter values used for data generation are provided in the Web materials. The vector of complete confounders $\bm{C}$ consisted of $C_1-C_4$, and the vector of incomplete confounders $\bm{Z}$ consisted of $Z_1$ and $Z_2$ as defined in Table \ref{tab: table1}. We generated $U, \bm{C}, \bm{Z}$ and $X$ sequentially, and fixed the exposure prevalence to around 50\% by modifying the intercept when generating the exposure. The auxiliary variable $A_1$ was generated to reflect, academic proficiency, a continuous measure in the VAHCS, and the mediator auxiliary variable $A_2$, sleep problems, which was binary: see Table \ref{tab: aux-simu} for the models used for generating these variables. We generated the outcome using a normal linear model with the following mean specification: 
\begin{equation}\label{equ: A-mod multi}
\begin{aligned}
E[Y|X,A_k,\bm{C},\bm{Z}]&=\theta_0+\theta_1 X+\theta_2C_1+\theta_3C_2+\theta_4C_3\\
&+\theta_5C_4+\theta_6Z_1+\theta_7Z_2 - 2\theta_1Z_2X + \theta_9A_k,
\end{aligned}
\end{equation}
where $k=1$ for m-DAG III and $k=2$ for m-DAG IV. We modified the strength of the main effect $\theta_1$ such that the true value of the ACE was a 0.3-unit difference in mean across all scenarios, this true value was set in order to achieve approximately 80\% power in the sample size of 1000 observations. The true value is indicated by a dashed line in Figure \ref{fig: multi-res}. We generated the unmeasured common cause for missingness indicators $W$ from a standard Normal distribution, then generated the missingness indicators sequentially, following the m-DAGs in Figure \ref{fig: multi-DAG} -- see Web materials for full details of the data generation models. The missingness proportions were fixed at 15\% for each incomplete confounder, and 20\% for the exposure and outcome, by modifying the intercept. The complete-case proportion was around 55\% across all m-DAGs. 

For each scenario described in this section, we generated 2000 datasets, which ensures that the Monte Carlo standard error of an expected 95\% coverage probability is below 0.5\%. 

\subsection{Target analysis}\label{subsec: analysis}
Except for the A-CCA approach, the target analysis for estimating the ACE was g-computation using a C-adjusted regression model with correctly specified exposure-confounder interactions, as follows:

Univariable missingness, no interaction scenario
\begin{equation}\label{equ: C-mod target 1}
E[Y|X,C]=\gamma_0+\gamma_1 X + \gamma_2C
\end{equation}

Univariable missingness, interaction scenario
\begin{equation}\label{equ: C-mod target 2}
E[Y|X,C]=\zeta_0+\zeta_1 X + \zeta_2C+\zeta_3C X  \end{equation}

Multivariable missingness scenario
\begin{equation}\label{equ: C-mod target 3}
E[Y|X,\bm{C},\bm{Z}]=\eta_0+\eta_1 X + \eta_2C_1+\eta_3C_2+\eta_4C_3+\eta_5C_4+\eta_6Z_1+\eta_7Z_2+\eta_8Z_2X
\end{equation}

With A-CCA, the target analysis was g-computation using correctly specified A-adjusted models, which expanded models (\ref{equ: C-mod target 1}-\ref{equ: C-mod target 3}) by including a main effect for the auxiliary variable. 

In CCA, A-CCA, or within an imputed dataset, the standard error for the ACE was estimated using the bootstrap with 200 draws. 

\subsection{Implementation of missing data methods}\label{subsec: implementation}
We evaluated the performance of the following non-MI and MI-based missing data methods, introduced in Section \ref{subsec: Missing data methods}. 

\paragraph{The CCA approach} It was evaluated as one of the most commonly used methods for handling missing data. Here, the ACE was estimated using g-computation using the correctly specified C-adjusted outcome models (\ref{equ: C-mod target 1}-\ref{equ: C-mod target 3}) applied to the complete cases.

\paragraph{The A-CCA approach} It was consisted in applying g-computation using correctly specified A-adjusted models to the complete cases. 

\paragraph{MI methods with auxiliary variables} We considered the MI methods introduced in Section \ref{subsec: Missing data methods}, where the auxiliary variables required to block biasing paths were incorporated in the imputation procedure as described below. After imputation, the ACE was estimated in each imputed dataset using g-computation using a C-adjusted model and bootstrap for SE estimation \cite{bartlett2020bootstrap}, and the results were pooled across imputations using Rubin’s rules \cite{rubin2004multiple}. 

\textit{FCS methods} We considered three FCS-based MI approaches, using REG, PMM, or CART. For all approaches, the auxiliary variable was incorporated as a predictor in the imputation model together with all variables included in the substantive analysis model. To accommodate exposure-confounder interactions, if present in the C-adjusted model (\ref{equ: C-mod target 2}-\ref{equ: C-mod target 3}), we incorporated two-way interactions as predictors in the imputation models for the REG and PMM approaches. Taking the simulation scenario in m-DAG III(a) as an example, the univariate imputation model for $X$ included $Z_2Y$, the univariate imputation model for $Z_2$ included $XY$, and the univariate imputation model for all other variables included $Z_2X$.  For the parametric imputation method REG, we used linear and logistic regression to impute continuous and binary variables, respectively. 

\textit{SMCFCS methods} The auxiliary variables were incorporated in SMCFCS using the two approaches described in Section \ref{subsec: Missing data methods}. In the first approach (A-SMCFCS), the A-adjusted outcome model was used as the target model. Therefore, the auxiliary variables were automatically used as imputation predictors for all incomplete variables, and the imputation was semi-compatible with the substantive analysis for estimating the ACE (which used the C-adjusted model). In the second approach (SMCFCS-manual), the C-adjusted outcome model was used as the target model. For the predictor matrix for non-outcome variables, auxiliary variables were manually included in the proposal distribution for imputing variables whose missingness was caused by them. Therefore, the SMCFCS-manual approach incorporated information from auxiliary variables in univariate imputation models, which again resulted in the overall imputation model being semi-compatible with the substantive analysis model. For comparison purposes, we also implemented the default SMCFCS approach, which uses a C-adjusted model as the target model for imputation, without using auxiliary variables in the imputation.

All analyses were carried out in R version [4.1.2] \cite{Rversion}. The FCS-based approaches were implemented using the {\tt{mice}} command \cite{van2011mice}, where {\tt{rpart}} \cite{therneau2015rpart} and {\tt{pmm}} \cite{van2011mice} methods were used to conduct PMM and CART. The SMCFCS-based approaches were implemented using the {\tt{smcfcs}} package \cite{bartlett2015smcfcs}. Each MI approach generated 20 imputed datasets, each based on 10 iterations, which appeared to provide good convergence, based on diagnostic plots. 

We obtained the following performance indicators for each method across simulations \cite{morris2019using}: the relative bias (RB) of the point estimate of the ACE compared to the true value (as a percent, \%), the empirical SE of the estimated ACE, the average of the estimated SE (Model-based SE), the mean square error (MSE), and the bias-eliminated (BE) coverage. The Monte Carlo SEs for all measures were also obtained; see the Supplementary Material for formulae. 

\subsection{Simulation results}\label{subsec: Results}
\begin{figure}[ht]
\centering
\includegraphics[width=\linewidth]{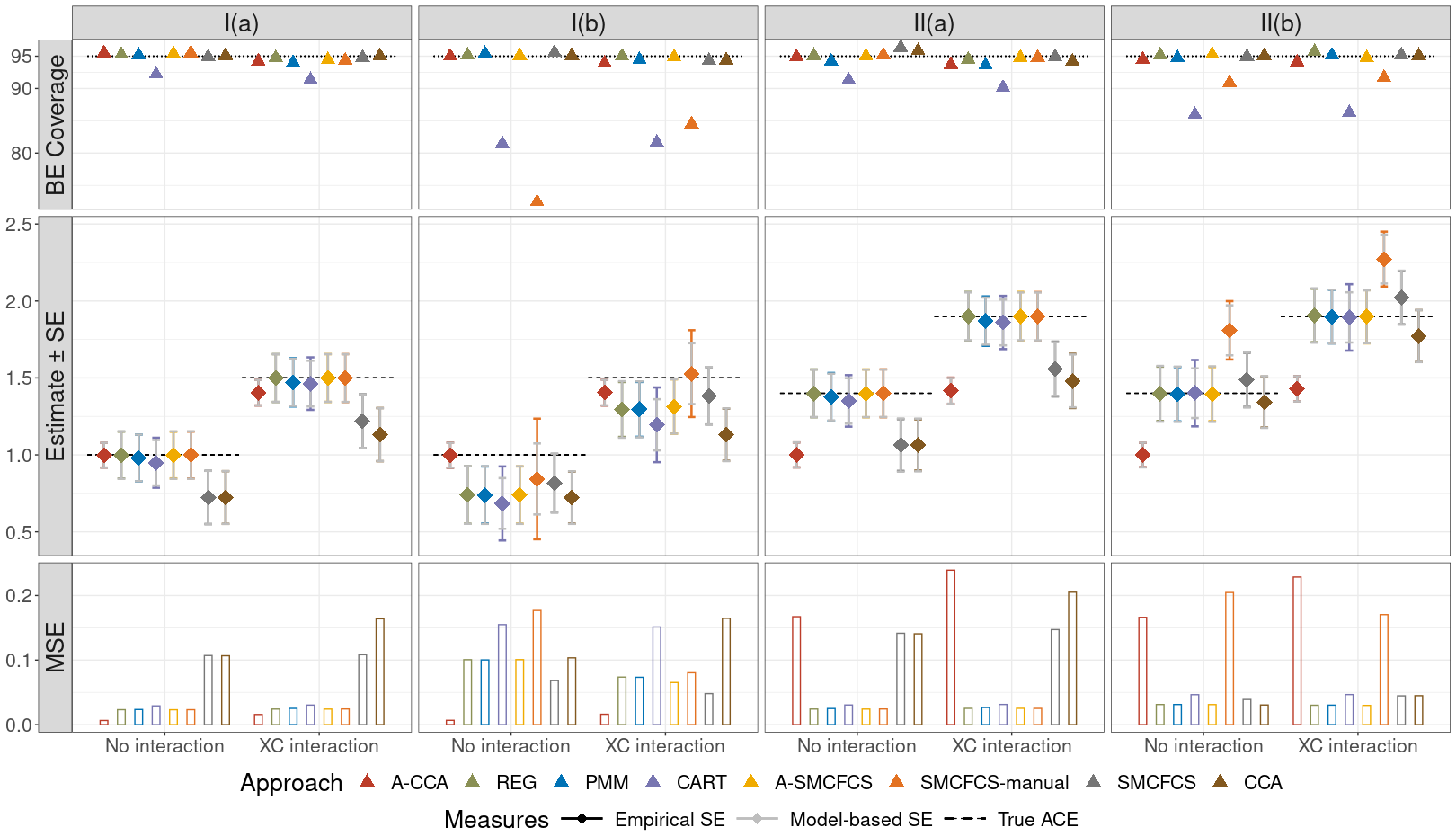}
\caption{Simulation results for missing data methods in univariable missingness mechanisms across the m-DAGs (columns) shown in Figure \ref{fig: uni-DAG} under outcome scenarios with and without exposure-confounder interaction (x-axes). Auxiliary variables were continuous. Panels show bias-eliminated coverage (top, dotted line = 95\%), mean of the ACE estimates with empirical and model-based SEs (middle, dashed lines represent the true values of the ACE), and mean squared error (bottom).}
\label{fig: uni-res}
\end{figure}

\paragraph{Univariable missingness mechanisms with strong associations} Figure \ref{fig: uni-res} presents the performance of missing data methods under univariable missingness mechanisms depicted by the m-DAGs in Figure \ref{fig: uni-DAG}, in settings with a continuous auxiliary variable. 
When only the outcome was incomplete and the auxiliary variable was not a mediator (m-DAG I(a)), both A-CCA and all MI approaches that included the auxiliary variables (i.e., all MI approaches except SMCFCS) were approximately unbiased (|RB|<10\%). However, when the auxiliary variable was a mediator (m-DAG II (a)), A-CCA showed substantial bias (|RB| > 20\%). Meanwhile, all MI approaches that incorporated the auxiliary variable remained approximately unbiased. 

When only the exposure was incomplete and the auxiliary variable was not a mediator (m-DAG I(b)), A-CCA was unbiased or approximately unbiased, whereas all MI approaches were biased except for SMCFCS-manual in the interaction scenario. In contrast, when the auxiliary variable was a mediator (m-DAG II(b)), only MI approaches that incorporated the auxiliary variable, except for SMCFCS-manual, were approximately unbiased.

Results from the binary auxiliary variable setting are presented in the Web materials and were generally similar to those observed in the continuous auxiliary variable setting. 
In summary, these results under simplified settings with strong associations revealed that MI methods that incorporate auxiliary variables (whether mediators or not), although inexact, only yielded clear bias in the specific scenario of m-DAG I(b) where the biasing path had the specific form of an open collider path between exposure and outcome. Meanwhile, as expected, A-CCA was clearly biased when the auxiliary variable was a mediator. Also as expected, methods not incorporating auxiliary variables at all (CCA and SMCFCS) were biased.

\begin{figure}[ht]
\centering
\includegraphics[width=\linewidth]{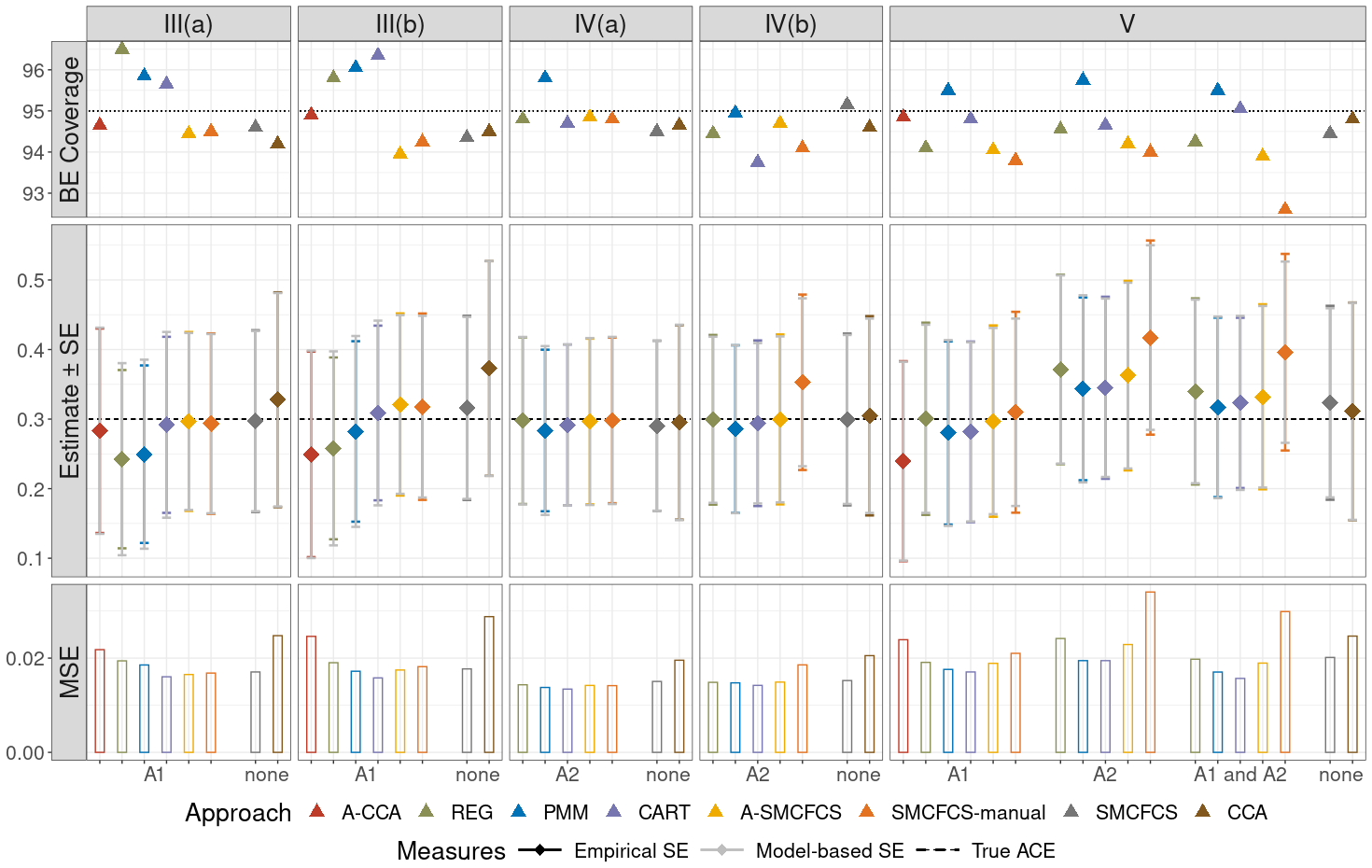}
\caption{Simulation results for missing data methods in multivariable missingness mechanisms across the m-DAGs (columns) shown in Figure \ref{fig: multi-DAG} in settings with the use of auxiliary variables (x-axes). Panels show bias-eliminated coverage (top, dotted line = 95\%), mean of the ACE estimates with empirical and model-based SEs (middle, dashed lines represent the true values of the ACE), and mean squared error (bottom). In m-DAGs IV(a), IV(b) and V, results from A-CCA using $A_2$ alone or both $A_1$ and $A_2$ are omitted from the plot, as they were biased by more than 50\%.}
\label{fig: multi-res}
\end{figure}

\paragraph{Multivariable missingness mechanisms with realistic associations} Figure \ref{fig: multi-res} presents the performance of missing data methods under multivariable missingness mechanisms depicted by the m-DAGs in Figure \ref{fig: multi-DAG}. Results from A-CCA using $A_2$ alone (m-DAGs IV(a), IV(b) V) or both $A_1$ and $A_2$ (m-DAG V) are omitted from the plot, as they were biased by more than 50\%.
When there was only one auxiliary variable $A_1$, which was a cause of outcome missingness and not a mediator (m-DAG III(a)), A-CCA, all implementations of SMCFCS, and CART, were approximately unbiased with close-to-nominal BE coverage, whereas PMM, REG and CCA were biased. When the auxiliary variable was a cause of exposure missingness and not a mediator (m-DAG III(b)), all implementations of SMCFCS, PMM and CART were approximately unbiased, although the latter two approaches showed important overcoverage. Other approaches exhibited bias. When the auxiliary variable was a mediator and predicted missingness in the outcome (m-DAG IV(a)), all MI approaches and CCA were approximately unbiased with all also showing appropriate BE coverage except for PMM which showed overcoverage. When the auxiliary variable was a mediator and predicted missingness in the exposure (m-DAG IV(b)), all MI approaches, except SMCFCS-manual, and CCA were approximately unbiased with close-to-nominal BE coverage except for CART, which showed undercoverage.

In m-DAG V when there were two mediators, all MI methods that used only $A_1$ in the imputation, as well as CART and PMM using both $A_1$ and $A_2$, and SMCFCS and CCA (which did not use any auxiliary variable) were approximately unbiased (|RB|<10\%). These methods also achieved close to 95\% BE coverage, except for REG, A-SMCFCS, and SMCFCS-manual using only $A_1$, which all showed undercoverage. Other methods were biased. 

In summary, these results under multivariable missingness settings with realistic associations results revealed that three methods were consistently approximately unbiased, with comparable precision: A-SMCFCS, CART and SMCFCS. While A-SMCFCS and SMCFCS methods showed slight undercoverage, CART showed slight overcoverage in some scenarios. Other methods showed variable performance in terms of bias across scenarios.

\section{Application to the VAHCS case study}\label{sec: Case}
We implemented the methods described in Section \ref{subsec: analysis} to the VAHCS case study, assuming the missingness mechanism depicted by m-DAG V in Figure \ref{fig: multi-DAG}. In this setting, academic proficiency ($A_1$) and sleep problems ($A_2$) were non-mediator and mediator auxiliary variables, respectively. Participant age ($U$), which was observed in the case study, was included in the imputation models as an auxiliary variable that may improve precision but is not required for recoverability. The target analysis for estimating the ACE was g-computation using outcome regression model (\ref{equ: C-mod target 3}), with the exposure-confounder interaction term and without auxiliary variables. We implemented the missing data methods both without auxiliary variables and with one or both of $A_1$ and $A_2$, always including $U$. For the MI approaches, missing values in auxiliary variables were imputed using logistic or linear regression univariate imputation model in MI methods other than PMM and CART. Each MI approach generated 40 imputed datasets based on 10 iterations, chosen according to the proportion of missingness and confirmed by convergence diagnostics.

Table \ref{tab:case-res} presents the results of the case study. Estimates from all methods suggest a moderate detrimental effect of adolescent cannabis use on young adulthood mental health among females. The estimates are similar across methods, ranging from 0.22 to 0.32, with standard errors of approximately 0.14, and with largely overlapping 95\% confidence intervals. 

With respect to the use of auxiliary variables, with FCS methods (REG, PMM, and CART) results obtained using $A_1$ only were very similar to those obtained without auxiliary variables or with both $A_1$ and $A_2$. In contrast, results using $A_2$ only showed slightly greater differences, except for REG. Estimates from REG and SMCFCS-based methods were generally larger than those from other approaches. Moreover, estimates from A-SMCFCS were more consistent across different auxiliary variable specifications compared with those from SMCFCS-manual.

\section{Discussion}\label{sec: Discussion}
In this paper, we investigated a set of m-DAGs that represent typical missingness mechanisms with auxiliary variables in epidemiological studies. For each of these m-DAGs, we derived recoverability expressions for the ACE and then investigated estimation methods in a simulation study. The latter evaluated implementations of MI that differed with respect to the inclusion of auxiliary variables in the imputation model and the extent of compatibility between the imputation and outcome models. 

In our investigation in simplified univariable missingness mechanisms with parameter settings pushed to extreme values, we found that MI methods that incorporate auxiliary variables were approximately unbiased except in the specific setting where the auxiliary variable is a common cause of the outcome and missingness in the exposure, and the exposure is a cause of its own missingness. Meanwhile, although a method using outcome regression with adjustment for auxiliary variables (A-CCA) was unbiased in the expected setting of a non-mediator variable and no interactions (see below), this approach was substantially biased when the auxiliary variable was a mediator and, to a lesser extent, in a setting with exposure-confounder interactions. Also, as expected, methods not incorporating auxiliary variables at all (CCA and SMCFCS) were biased. 

Under multivariable missingness settings with realistic associations results, we found that three methods were consistently approximately unbiased, with comparable precision: 1. the SMCFCS without using auxiliary variables, 2. the SMCFCS method that is compatible with an outcome regression adjusted for all non-mediator auxiliary variables (A-SMCFCS), and 3. the FCS-based MI using regression trees (CART) that incorporated non-mediator auxiliary variables. This is despite the presence of problematic paths like the one identified in the univariable setting in some of the multivariable m-DAGs examined. While A-SMCFCS and SMCFCS methods showed slight undercoverage, CART showed slight overcoverage in some scenarios. Other methods showed variable performance in terms of bias across scenarios. 

In terms of implications for practice, taken together these findings suggest that in realistic settings, MI methods incorporating auxiliary variables, specifically A-SMCFCS and CART, are the preferred approaches. Additionally, while A-CCA is easy to apply and unbiased in certain scenarios, it can be extremely biased and thus should be avoided in the context of mediator auxiliary variables.  

Discussion of how auxiliary variables can be used to identify causal effects can be traced back to earlier work by Correa et al. \cite{correa2018generalized} and further extended by Mathur et al. \cite{mathur2025common,mathur2024imputation}. The latter proposed a criterion for determining the recoverability of the conditional average treatment effect (CATE) when the auxiliary variable is not a mediator. Specifically, they introduced the concept of a sufficient adjustment set (SAS) \cite{mathur2025common,mathur2024imputation}, which includes confounders and auxiliary variables that block all backdoor paths between the outcome and the observation indicator (i.e., an indicator of whether the record is complete). They showed that the CATE, where the conditioning is with respect to the SAS variables, is recoverable and can be unbiasedly estimated from the fully observed records using outcome regression adjusting for the SAS variables, i.e. the A-CCA approach. Our results are consistent with their findings, which apply to settings where the auxiliary variable is not a mediator. Indeed, when we consider m-DAGs I(a) and I(b) in scenarios with no interactions, so that the causal effect is constant across strata defined by the SAS variables and the ACE and CATE are identical, we find that the A-CCA approach is unbiased.

Including auxiliary variables in the imputation is widely adopted in missing data literature, where the general recommendation is to include ``useful’’ auxiliary variables that are at least correlated with the incomplete variable(s) for improving efficiency, and that may also reduce bias if they also predict the missingness \cite{carpenter2023multiple,mainzer2024comparison}. In the current paper, using the perspective of a formal causal inference approach considering the recoverability (identifiability) of the ACE, we show that, for the task of estimating the ACE, the inclusion of auxiliary variables can introduce bias depending on the m-DAG and imputation approach. Specifically, regression-based (REG) and predictive mean matching (PMM) FCS were biased in our simulations under multivariable missingness mechanisms when including an auxiliary variable that was not a mediator, whereas in these settings the SMCFCS approach without using an auxiliary variable was unbiased. In addition, using an auxiliary variable in SMCFCS inappropriately also gave rise to bias, which happened when a mediator auxiliary variable was manually specified as an imputation predictor while using the correct outcome model (i.e. the C-adjusted model) in the substantive analysis (see results for m-DAGs II (b) and IV (b)). By contrast, CART, a non-parametric imputation approach, was approximately unbiased across the majority of multivariable scenarios we evaluated. However, in some univariable missingness settings, CART estimates were biased and showed undercoverage.

A strength of our research is that we investigated recoverability and estimation of the ACE in a series of m-DAGs with auxiliary variables that depict typical missingness mechanisms covering a wide range of scenarios in observational studies. These considered two distinct types of auxiliary variables: mediators and non-mediators of the exposure-outcome relationship. Findings relating to the mediator type are particularly useful for longitudinal studies with repeated measures across waves of follow-up, where earlier measures of the outcome are highly plausible candidate auxiliary variables. In the simulation studies, we evaluated MI-based methods that are compatible with the presence of exposure-confounder interactions in the outcome model. This is to ensure that the performance of the methods reflected their ability to incorporate additional variables, like the auxiliary variables, which represented the source of incompatibility of focus here. 

A limitation of our research is the inevitably limited range of scenarios examined, meaning findings may not generalise to other settings. In particular, the associations examined in the multivariable setting may not be reflective of associations in other settings, and we cannot rule out potential bias in methods that were approximately unbiased here. For example, our findings in the univariable setting with strong associations revealed a specific causal structure where MI methods exhibit more bias (m-DAG I(b)), but this was minimal in realistic scenarios with multivariable missingness. This could be due to the weaker associations but also due to the multivariable nature requiring multiple models. Future research further examining the theoretical basis for this is warranted. Additionally, we did not consider the common scenario where there is missing data in the auxiliary variables in question. This is an open and important question not addressed in this paper \cite{carpenter2023multiple,mathur2025pitfalls}. Establishing recoverability results in such a setting is more complex as the variety of possible causal structures increases. Using incomplete auxiliary variables in MI also requires a strategy to impute them, which can make specification of parametric models or applications of SMCFCS particularly challenging \cite{carpenter2023multiple}. The non-parametric imputation methods may therefore provide a promising direction for future research in this context. 

Another related topic that was out of scope here but is critical for the field is the consideration of time-varying exposures and outcomes. This is a common setting in observational studies, and the role of auxiliary variables in such longitudinal settings deserves careful study \cite{holovchak2025recoverability}. Also, alternative estimation methods for the ACE are worth further investigation, particularly in light of recent developments in doubly robust estimation \cite{dashti2024handling,seaman2018introduction}, which offer a promising avenue for tackling causal inference and missing data problems. 

In summary, we investigated the role of auxiliary variables in the recoverability and estimation of the ACE under several typical missingness mechanisms depicted by m-DAGs. Our findings have important implications for practice, suggesting that use of compatible (A-SMCFCS) and flexible (CART) MI methods that incorporate auxiliary variables provide a good solution. Meanwhile other methods, especially those that do not include such variables (MI-based or not, like CCA) or methods that may incorporate mediator auxiliary variables inappropriately (A-CCA) may lead to substantial bias and should be avoided.

\section{DECLARATIONS}

\noindent {\bf{Supplementary material}}\\
Supplementary material is available at Statistics in Medicine Journal online.
\\
\noindent {\bf{Ethics approval}}\\
The case study used data from the Victorian Adolescent Health Cohort Study. Data collection protocols were approved by The Royal Children’s Hospital’s Ethics in Human Research Committee. Informed parental consent was obtained for each participant prior to entry.
\\
\noindent {\bf{Funding}}\\
The following authors were supported by National Health and Medical Research Council (NHMRC) Investigator Grants: SGD (ID 2027171), KJL (ID 2017498), MMB (ID 2009572). The Murdoch Children’s Research Institute is supported by the Victorian Government’s Operational Infrastructure Support Program. The funding bodies did not have any role in the study design, analysis, interpretation or writing of the paper.
\\
\noindent {\bf{Data availability}}\\
The data underlying this article will be shared on reasonable request to the corresponding author with the permission of the Chief Investigators of the study, Professor Craig Olsson and Professor Susan Sawyer.\\
\noindent {\bf{Acknowledgements}}\\
The authors would like to thank the Victorian Centre for Biostatistics (ViCBiostat), Causal Inference group, Missing Data group and other members of ViCBiostat for providing feedback in designing and interpreting the simulation study. We also wish to thank the families who participated in the VAHCS, the study research team, and the Principal Investigator of the study, the late Professor George Patton.
\\
\noindent {\bf{Author contributions}}\\
JZ, SGD, JBC, KJL, and MMB conceived the project and designed the study. JZ completed the coding and designed the simulation study, with input from co-authors, and drafted the manuscript. MMB, SGD, JBC, and KJL provided critical input to the manuscript. All of the co-authors read and approved the final version of this paper.
\\
\noindent {\bf{Conflict of interest}}\\
None declared


\begin{landscape}
\begin{table}[]
\caption{Descriptive statistics for variables in the motivating example, using data from the Victorian Adolescent Health Cohort Study ($n$= 1000)}
\label{tab: table1}
\resizebox{\columnwidth}{!}{%
\begin{tabular}{lllcccc}
\hline
\multirow{2}{*}{} & \multirow{2}{*}{Variable} & \multirow{2}{*}{Type} & \multirow{2}{*}{Notation} & \multicolumn{2}{c}{Stratified by exposure $n$ (\%) or mean (SD)\textsuperscript{a}} & \multirow{2}{*}{Missing (\%)} \\
 &  &  &  & Exposed & Unexposed &  \\ \hline
\multirow{2}{*}{\begin{tabular}[c]{@{}l@{}}Auxiliary \\ variables\end{tabular}} & Academic proficiency & \begin{tabular}[c]{@{}l@{}} Self-rated academic proficiency $z$-score\textsuperscript{b}, \\ average across waves 2 to 6\end{tabular} & $A_1$ & -0.11 (0.93) & 0.42 (0.95) & 0.2 \\ \cline{2-7} 
 & Sleep problems  & \begin{tabular}[c]{@{}l@{}}
 Self-reported sleep problems\\ 0= No at wave 6\\ 1= Yes at wave 6 \end{tabular} & $A_2$ & 205 (33.8) & 42 (58.3) & 15.2 \\ \hline
\multirow{4}{*}{\begin{tabular}[c]{@{}l@{}}Complete \\ confounders\end{tabular}} & Parental education & \begin{tabular}[c]{@{}l@{}}Parents complete high-school education\\ 0=Complete by wave 2\\ 1=Failure by wave 2\end{tabular} & $C_1$ & 206 (34.2) & 35 (41.7) & 3.4 \\ \cline{2-7} 
 & Parental divorce & \begin{tabular}[c]{@{}l@{}}Parental divorce or separation\\ 0=No by wave 6\\ 1=Yes by wave 6\end{tabular} & $C_2$ & 95 (15.7) & 39 (45.3) & 0.1 \\ \cline{2-7} 
 & Antisocial behaviour & \begin{tabular}[c]{@{}l@{}}Self-reports on any of theft, interpersonal violence \\ and property damage activities in adolescence waves\\ 0= No across all waves 2 to 6\\ 1= Yes in any waves of 2 to 6\end{tabular} & $C_3$ & 43 (7.1) & 33 (38.4) & 0.6 \\ \cline{2-7} 
 & Adolescent smoking & \begin{tabular}[c]{@{}l@{}}Self-reported daily smoking in adolescence waves\\ 0= No across all waves 2 to 6\\ 1= Yes in any waves of 2 to 6\end{tabular} & $C_4$ & 101 (16.7) & 63 (73.3) & 0 \\ \hline
 \multirow{2}{*}{\begin{tabular}[c]{@{}l@{}}Incomplete \\ confounders\end{tabular}} & Alcohol use & \begin{tabular}[c]{@{}l@{}}Self-reported a frequency of drinking of more than \\ three days a week in adolescence waves\\ 0= No across all waves 2 to 6\\ 1= Yes in any waves of 2 to 6\end{tabular} & $Z_1$ & 156 (25.7) & 67 (88.2) & 21.0 \\ \cline{2-7} 
 & Adolescent depression & \begin{tabular}[c]{@{}l@{}}Participant recorded a total CIS-R score of 12 or \\ higher in adolescence waves\\ 0= No across all waves 2 to 6\\ 1= Yes in any waves of 2 to 6\end{tabular} & $Z_2$ & 300 (49.5) & 62 (81.6) & 13.8 \\ \hline
Exposure & Cannabis use & \begin{tabular}[c]{@{}l@{}}Self-reported the use of cannabis more than once \\ a week in adolescence waves\\ 0= No across all waves 2 to 6\\ 1= Yes in any waves of 2 to 6\end{tabular} & $X$ & 606 (87.6) & 86 (12.4) & 30.8 \\ \hline
Outcome & Adulthood mental health & The computerised CIS-R $z$-score\textsuperscript{b} at wave 7 & $Y$ & -0.12 (1.00) & 0.48 (0.87) & 13.4 \\ \hline
\begin{tabular}[c]{@{}l@{}}Common cause \\of $X,\bm{C},\bm{Z}$ \end{tabular} & Participant's age & Participant's age ($z$-score\textsuperscript{b}) at wave 2 & $U$ & -0.12 (0.80) & 0.11 (0.93) & 9.3 \\ \hline
\multicolumn{7}{l}{\begin{tabular}[c]{@{}l@{}} Abbreviations: CIS Clinical Interview Schedule. \\ a. For incomplete variables, the descriptive statistics are reported among those with observed data for the variable.\\ b. In standard deviation units, standardised to the overall sample.
\end{tabular}}
\end{tabular}%
}
\end{table}
\end{landscape}

\begin{table}[]
\caption{Recoverability results for each m-DAG in Figures \ref{fig: uni-DAG} and \ref{fig: multi-DAG}}
\label{tab: recoverability}
\resizebox{\columnwidth}{!}{%
\begin{tabular}{cc}
\hline
m-DAG & Recoverability expression for $P(y^x)$ \\ \hline
I (a) & $\displaystyle \sum_{a_1,\bm{c}}P(y|x,a_1,\bm{c},M_Y=0)P(a_1,\bm{c})$ \\
I (b) & $\displaystyle \sum_{a_1,\bm{c}}P(y|x,a_1,\bm{c},M_X=0)P(a_1,\bm{c})$ \\
II (a) & $\displaystyle \sum_{a_2,\bm{c}}P(y|x,a_2,\bm{c},M_Y=0)P(a_2|x,\bm{c})P(\bm{c})$ \\
II (b) & $\displaystyle \sum_{a_2,\bm{c}}\frac{P(y,x|a_2,\bm{c},M_X=0) P(a_2,\bm{c})}{\sum_{a_2}P(x|a_2,\bm{c},M_X=0) P(a_2)}$ \\ \hline
III (a) & $\displaystyle \sum_{a_1,\bm{c},\bm{z}} P(y|x,a_1,\bm{c},\bm{z},M_{all}=0) P(a_1,\bm{c},\bm{z}|M_{\bm{Z}}=0) \frac{P(M_{\bm{Z}}=0)}{P(M_{\bm{Z}}=0|a_1,\bm{c})}$ \\
III (b) & $\displaystyle \sum_{a_1,\bm{c},\bm{z}} P(y|x,a_1,\bm{c},\bm{z},M_{all}=0) P(a_1,\bm{c},\bm{z}|M_{\bm{Z}}=0) \frac{P(M_{\bm{Z}}=0)}{P(M_{\bm{Z}}=0|a_1,\bm{c})}$ \\
IV (a) & $\displaystyle \sum_{a_2,\bm{c},\bm{z}} P(y,\bm{z}|x,a_2,\bm{c},M_{all}=0) P(a_2|x,\bm{c},M_X=0) P(\bm{c})$ \\
IV (b) & $\displaystyle \sum_{a_2,\bm{c},\bm{z}} \frac{P(y,x,\bm{z}|a_2,\bm{c},M_{all}=0) P(a_2,\bm{c})} {\sum_{a_2}P(x|a_2,\bm{c},M_X=0) P(a_2)}$ \\
V &  $\begin{aligned}\sum_{a_1,a_2,\bm{c},\bm{z}} & P(y|x,a_1,a_2,\bm{c},\bm{z},M_{all}=0) P(a_2|x,a_1,\bm{c},\bm{z},M_X=M_{\bm{Z}} = 0) \times \\
&\frac{P(\bm{c},\bm{z},M_{\bm{Z}} = 0) P(M_Y = 0,a_1|a_2,\bm{c},\bm{z},M_{\bm{Z}} = 0)}{P(M_Y = M_{\bm{Z}} = 0|a_1,a_2,\bm{c})} \end{aligned}$ \\
\hline

\end{tabular}%
}
\end{table}

\begin{table}[]
\caption{Generation models for auxiliary variables in  simulation study}
\label{tab: aux-simu}
\resizebox{\columnwidth}{!}{%
\begin{tabular}{clll}
\hline
 & \multicolumn{3}{c}{Auxiliary variable generation models} \\ \hline
\multicolumn{1}{c|}{m-DAG I} & \begin{tabular}[c]{@{}l@{}}Binary auxiliary variable setting:\\ $A_1 \sim Bern(\text{expit}(-0.2+C))$\end{tabular} & \multicolumn{2}{l}{\begin{tabular}[c]{@{}l@{}}Continuous auxiliary variable setting:\\ $A_1 \sim N(0.3C, 1)$\end{tabular}} \\ \hline
\multicolumn{1}{c|}{m-DAG II} & \begin{tabular}[c]{@{}l@{}}Binary auxiliary variable setting:\\ $A_2 \sim Bern(\text{expit}(-0.2+X))$\end{tabular} & \multicolumn{2}{l}{\begin{tabular}[c]{@{}l@{}}Continuous auxiliary variable setting:\\ $A_2 \sim N(0.3X, 1)$\end{tabular}} \\ \hline
\multicolumn{1}{c|}{m-DAG III} & \multicolumn{3}{l}{$A_1 \sim N(-0.27+0.28 C_1+0.08 C_2+0.25 C_3+0.46 C_4,1)$} \\ \hline
\multicolumn{1}{c|}{m-DAG IV} & \multicolumn{3}{l}{$A_2 \sim Bern(\text{expit}(-0.67+X))$} \\ \hline
\end{tabular}%
}
\end{table}

\begin{table}[]
\caption{Estimates of ACE obtained using missing data methods in the case study ($n$= 961)}
\label{tab:case-res}
\resizebox{\columnwidth}{!}{%
\begin{tabular}{cccc}
\hline
\multicolumn{1}{c|}{Approach} & Estimate & Standard error & 95\% Confidence interval \\ \hline
\multicolumn{4}{l}{Auxiliary variable used in imputation: no} \\ \hline
\multicolumn{1}{c|}{CCA} & 0.27 & 0.15 & (-0.04, 0.54) \\
\multicolumn{1}{c|}{PMM} & 0.25 & 0.13 & (-0.01, 0.51) \\
\multicolumn{1}{c|}{CART} & 0.23 & 0.13 & (-0.03, 0.49) \\
\multicolumn{1}{c|}{REG} & 0.30 & 0.13 & (0.03, 0.56) \\
\multicolumn{1}{c|}{SMCFCS} & 0.28 & 0.14 & (-0.01, 0.56) \\ \hline
\multicolumn{4}{l}{Auxiliary variable used in imputation: $A_1$} \\ \hline
\multicolumn{1}{c|}{A-CCA} &  0.27 &  0.15 & (-0.02, 0.58) \\
\multicolumn{1}{c|}{PMM} & 0.25 & 0.14 & (-0.03, 0.52) \\
\multicolumn{1}{c|}{CART} & 0.23 & 0.14 & (-0.05, 0.50) \\
\multicolumn{1}{c|}{REG} & 0.30 & 0.16 & (-0.01, 0.61) \\
\multicolumn{1}{c|}{SMCFCS-manual} & 0.30 & 0.14 & (0.03, 0.58) \\
\multicolumn{1}{c|}{A-SMCFCS} & 0.32 & 0.14 & (0.04, 0.59) \\ \hline
\multicolumn{4}{l}{Auxiliary variable used in imputation: $A_2$} \\ \hline
\multicolumn{1}{c|}{A-CCA} & 0.22 & 0.13 & (-0.03, 0.47) \\
\multicolumn{1}{c|}{PMM} & 0.22 & 0.14 & (-0.06, 0.51) \\
\multicolumn{1}{c|}{CART} & 0.27 & 0.13 & (0, 0.53) \\
\multicolumn{1}{c|}{REG} & 0.30 & 0.15 & (0.01, 0.59) \\
\multicolumn{1}{c|}{SMCFCS-manual} & 0.25 & 0.14 & (-0.02, 0.53) \\
\multicolumn{1}{c|}{A-SMCFCS} & 0.28 & 0.15 & (-0.02, 0.57) \\ \hline
\multicolumn{4}{l}{Auxiliary variable used in imputation: $A_1$ and $A_2$} \\ \hline
\multicolumn{1}{c|}{A-CCA} & 0.22 & 0.16 & (-0.07, 0.54) \\
\multicolumn{1}{c|}{PMM} & 0.24 & 0.16 & (-0.06, 0.55) \\
\multicolumn{1}{c|}{CART} & 0.22 & 0.14 & (-0.06, 0.5) \\
\multicolumn{1}{c|}{REG} & 0.30 & 0.14 & (0.02, 0.57) \\
\multicolumn{1}{c|}{SMCFCS-manual} & 0.32 & 0.14 & (0.05, 0.6) \\
\multicolumn{1}{c|}{A-SMCFCS} & 0.29 & 0.15 & (0, 0.58) \\ \hline
\end{tabular}%
}
\end{table}

\clearpage
\bibliographystyle{unsrtnat}  

\bibliography{article.bib}

\end{document}